\documentclass[twocolumn,superscriptaddress,showpacs,preprintnumbers,amsmath,amssymb]{revtex4-1}

\usepackage{amsmath}
\usepackage{amssymb}
\usepackage{graphicx}
\usepackage{morefloats}
\usepackage[usenames,dvipsnames]{xcolor}
\usepackage{blkarray}
\usepackage{verbatim}
\usepackage{hyperref}
\usepackage{color}

\begin{document}

\title{Conformism-driven phases of opinion formation on heterogeneous networks: \\the q-voter model case}

\author{Marco Alberto Javarone}
\affiliation{Department of Mathematics and Computer Science, University of Cagliari, Cagliari (Italy)}
\affiliation{DUMAS - Department of Humanities and Social Sciences, University of Sassari, Sassari (Italy)}

\author{Tiziano Squartini}
\affiliation{Institute for Complex Systems UOS Sapienza, ``Sapienza" University of Rome, P.le Aldo Moro 5, 00185 Rome (Italy)}

\date{\today}

\begin{abstract}
The q-voter model, a variant of the classic voter model, has been analyzed by several authors: while allowing to study opinion dynamics, this model is also believed to be one of the most representative among the many defined in the wide field of sociophysics. 
Here, we investigate the consequences of conformity on the consensus reaching process, by numerically simulating a q-voter model with agents behaving either as conformists or non-conformists, embedded on heterogeneous network topologies (as small-world and scale-free).
In fact, although it is already known that conformity enhances the reaching of consensus, the related process is often studied only on fully-connected networks, thus strongly limiting our full understanding of it. This paper represents a first step in the direction of analyzing more realistic social models, showing that different opinion formation-phases, driven by the conformist agents density, are observable. 
As a result, we identify threshold values of the density of conformist agents, varying across different topologies and separating different regimes of our system, ranging from a disordered phase, where different opinions coexist, to a gradually more ordered phase, where consensus is eventually reached.
\end{abstract}

\maketitle

Recent years have witnessed the increasing interest of scientists belonging to different fields, as physics, mathematics and computer science, for socio-economic systems~\cite{econo-socio,galam01,loreto01,perc01,javarone01}. In particular, it has become evident that several models born within the realm of statistical physics can be successfully employed for understanding simplified social systems, thus gaining insight in the human behavior~\cite{loreto01}. 

More precisely, the influence of conformity (considered a fundamental social trait) on opinion dynamics has been extensively studied~\cite{javarone02,galam02,galam03,galam04,intra}. Although conformity is of great interest in social sciences, e.g. in social psychology~\cite{aronson01}, several authors have analyzed it by adopting the viewpoint of statistical mechanics~\cite{weron04,weron05}. Just to cite a few, in~\cite{galam02,galam03} authors analyzed the role of contrarians (i.e. agents acting as non-conformists) in voting dynamics and in~\cite{javarone02,galam03} authors analyzed how conformity affects opinion dynamics by implementing the local majority rule.

In this work, we approach the problem of understanding how conformity affects opinion dynamics by implementing the q-voter model~\cite{loreto01,castellano01,galam05,timpanaro01,galam06}, i.e. a variant of the classic voter model~\cite{redner01}, on heterogeneous networks. In fact, while it is already known that conformity enhances the reaching of consensus (i.e. an opinion shared by \emph{all} agents)~\cite{javarone02} the details of this process are still questioned~\cite{timpanaro01}. Moreover, systems like the voter model and the q-voter model are often simulated over fully connected networks~\cite{castellano01,weron03,hisakado01}, and only to a lesser extent on more complex topologies (see for instance~\cite{vazquez01,sood01,satorras01}). If, on the one hand, this allows to analytically model the system under the mean-field approximation, on the other strongly limits the validity of results to unrealistic scenarios as it has been proven that social systems show highly heterogeneous structures~\cite{estrada01}. Thus, our analysis aims at exploring the behavior of the q-voter model by considering more realistic network topologies in order to understand the extent to which $1)$ varying the amount of conformist agents and $2)$ varying the network structure affects the consensus reaching process. In order to do so, we heavily rest upon numerical simulations. 

Results of our simulations indicate the presence of different opinion-formation regimes, driven by the density of conformist agents and varying across different network configurations. Threshold values separating different regimes vary as well. Moreover, the system seems to undergo a spontaneous symmetry-breaking, by (stochastically) choosing states with the same ``net'' opinion but opposite signs.
\newline
\newline
\indent In order to study the role of conformity in the q-voter model, we defined a simple agent-based model by considering $N$ agents, provided with an opinion and a social character.

Opinions are mapped to the agents states $s_i= \pm1,\:i=1\dots N$ and are assigned to each agent of the population stochastically, i.e. according to the probability coefficients $P_+=P_-=1/2$; thus, our initial expected number of opinions $+1$ is $\langle N_+^0\rangle=N/2$. Moreover, agents are provided with an individual behavior, i.e. either conformist or non-conformist. In what follows, we will adopt the definition according to which a conformist agent adopts the opinion of the majority of its neighbors, whereas a non-conformist one adopts the opposite.
As for the opinion, the behavior is assigned stochastically too, according to the coefficients $P_c$ and $P_a\equiv 1-P_c$, i.e. the probability to behave as a conformist or a non-conformist, respectively. As before, the initial expected number of conformist agents is $\langle N_c^0\rangle=P_c\cdot N$. 
The two processes of assigning opinions and behaviors are independent: so, each agent's initial probability of being both conformist and having opinion $+1$ is $p^0_{c,+}\equiv P_c\cdot P_+=\frac{P_c}{2}$. We will consider agents interacting on different configurations: while the probabilities $P_+$ and $P_-$ will remain fixed, $P_c$ and $P_a$ will vary, in order to achieve different densities of conformist (and non-conformist) agents in the population. Naturally, opinions vary as a result of the system dynamics. 

The q-voter model extends the classic voter model, letting each agent adopt the opinion shared by a subset of neighbors of arbitrary dimension~\cite{castellano01}. This model is described by two parameters: $q$ and $\epsilon$. The former represents the number of neighbors each agent has to consider to have its opinion defined, whereas the latter represents the probability for each agent to change its state anyway, even if not all the $q$ chosen neighbors agree. We implement the q-voter model setting $q=4$ and $\epsilon = 0$ (see Appendix \textbf{B} for different choices of $q$). Therefore, agents choose $q=4$ neighbors at random: if they all share the same opinion, a conformist agent adopts it, whereas a non-conformist agent adopts the opposite one. Otherwise, the agent keeps its precedent opinion: in fact, setting $\epsilon = 0$ means setting to zero the probability of changing opinion stocastichally, in the event the $q$ neighbors disagree.

It is worth emphasizing that the implemented updating rule has been chosen to be synchronous; this means that every agent updates its state simultaneously, on the basis of neighbors' opinion at the previous time step. In fact, we believe asynchronous updating does not adequately capture the real dynamics of a social experiment. 
For instance, let us imagine many people forming groups to discuss about politics: it is hard to imagine participants discussing and changing their opinion ``sequentially''. Persons interacts with their neighbors simultaneously, updating their opinion in ``real time'', i.e. \emph{before} being engaged in a new discussion with a different group. Another example is provided by voting scenarios, where people express their opinion at the same time. Moreover, even if asynchronous updating were applicable, it would cause the system dynamics to be dependent on the particular sequence of agents chosen.

\begin{figure}[t!]
\centering
\includegraphics[width=0.395\textwidth]{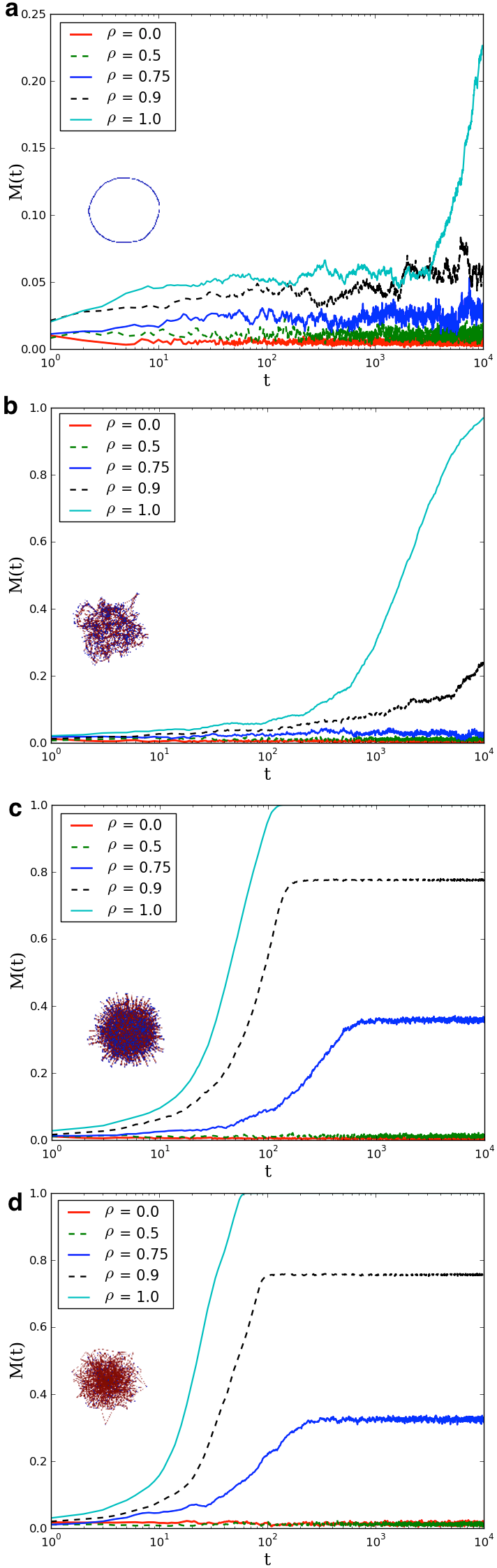}
\caption{\small Evolution of the magnetization for different values of $\rho$ and different configurations: \textbf{a} regular lattice; \textbf{b} small-world network ($\beta = 0.1$); \textbf{c} small-world network ($\beta = 0.5$); \textbf{d} scale-free network. Small pictorial representations are shown for each network.\label{fig:magnetization}}
\end{figure}
\begin{figure*}[t!]
\centering
\includegraphics[width=0.49\textwidth]{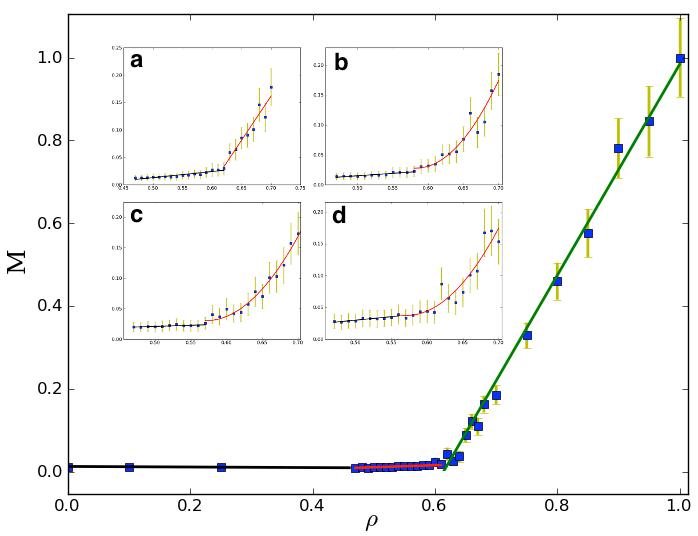}
\includegraphics[width=0.49\textwidth]{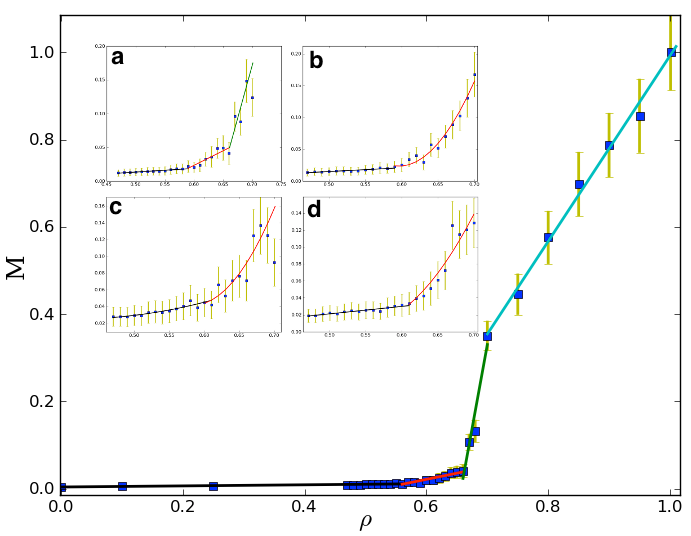}
\caption{Phase-diagram plotting $M$ versus $\rho$ for two network configurations (left: scale-free; right: small-world with $\beta=0.5)$: different phases are visible, separated by threshold values of $\rho$. Insets show the same analysis for networks with ({\bf a}) $N=2500$, ({\bf b}) $N=2000$, ({\bf c}) $N=1000$ and ({\bf d}) $N=500$. Error bars represent the standard deviation over the simulations run. The average $R^2$ of the fits is: scale-free - (main panel) 0.9, $({\bf a})$ 0.88, $({\bf b})$ 0.88, $({\bf c})$ 0.8, $({\bf d})$ 0.86; Watts-Strogatz - (main panel) 0.92, $({\bf a})$ 0.85, $({\bf b})$ 0.92, $({\bf c})$ 0.9, $({\bf d})$ 0.86. \label{fig:phase}}
\end{figure*}

Numerical simulations of our model have been carried on by chosing $N = 5000$ agents, embedded on different network topologies as scale-free networks, regular lattices, small world networks and completely random networks. While scale-free networks have been generated via the Barabasi-Albert model~\cite{barabasi01}, the last three kinds of networks have been generated via the Watts-Strogatz model~\cite{watts01}. The latter allows to obtain different network configurations by varying the value of the rewiring probability $\beta$: regular lattices are achieved by setting $\beta = 0$, small-world networks by setting $0<\beta<1$ and completely random networks by setting $\beta = 1$. In this work, we have considered the following values: $\beta \in [0, 0.01, 0.1, 0.5, 1]$. Moreover, all the considered networks have an average degree equal to $\sum_{i=1}^Nk_i/N=8$ (i.e. agents have, on average, eight neighbors).
Each simulation has been performed with a different amount of conformist agents, $\rho \in [0, 0.1, 0.25, 0.5, 0.6, 0.65, 0.7, 0.75, 0.9, 1]$, and it has been run for $10^4$ time steps. For the vast majority of cases this temporal limit was long enough to reach a steady-state, as only few network configurations required more time. However, in the latter scenarios (e.g. regular lattices) we performed longer simulations (see Appendix \textbf{A}).
\newline
\newline
\indent We first consider the evolution of the system magnetization over time, i.e. the absolute value of the difference between the number of agents in the two states~\cite{mobilia01}, normalized to $N$:
\begin{equation}\label{eq:magnetization}
M = \frac{|N_+ - N_-|}{N}.
\end{equation}

The magnetization ranges between 0 and 1 ($0\leq M\leq1$), with $M = 0$ indicating the equipartition of the two opinions (i.e. the maximally disordered phase), and $M = 1$ indicating that consensus has been reached. Notice that both situations $N_+=0,\:N_-=N$ and, vice-versa, $N_-=0,\:N_+=N$ are compatible with consensus, i.e. magnetization is uninformative about the dominant opinion sign. 

Figure~\ref{fig:magnetization} illustrates the evolution of the magnetization, upon varying the value of $\rho$ for different network configurations. Remarkably, the density of conformist agents strongly affects the process of consensus reaching; more detailedly, $1)$ values of $\rho\leq 0.5$ seem not to be sufficiently high to let the system escape the disordered phase where the two opinions coexist; $2)$ values of $0.5<\rho<1$ let the system escape the disordered phase but not to reach consensus: a steady-state is reached where one of the two opinions prevails on the other; $3)$ only the density value $\rho = 1$ allows the system to reach the consensus. 

\begin{figure*}[t!]
\centering
\includegraphics[width=0.99\textwidth]{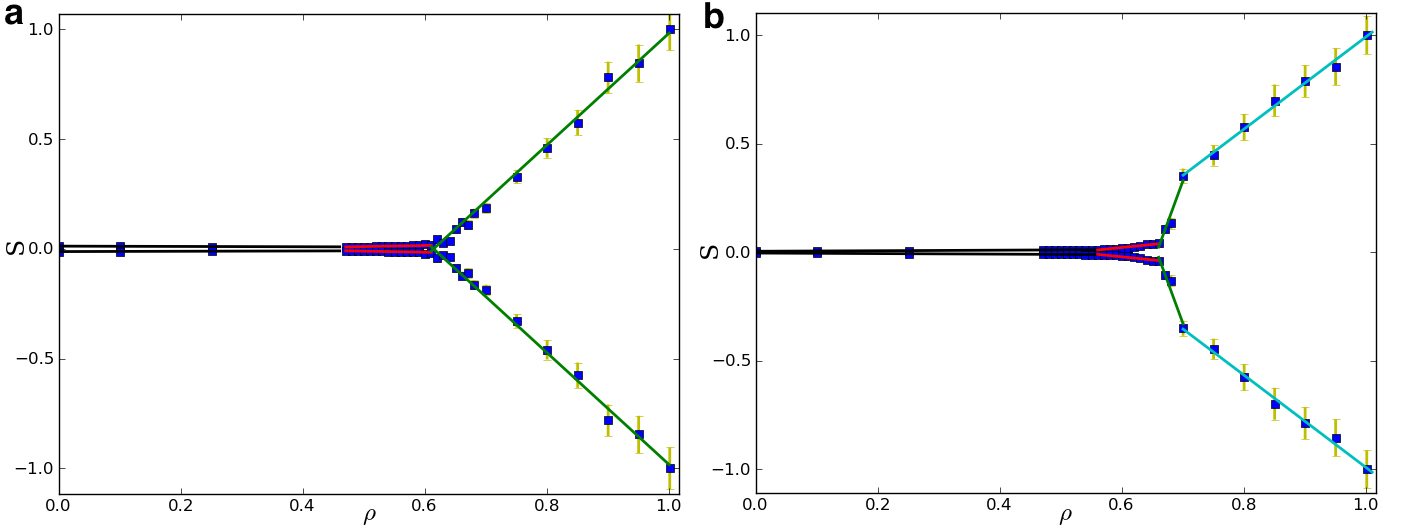}
\caption{Symmetry-breaking diagram plotting $S$ versus $\rho$ for two network configurations (left: scale-free; right: small-world with $\beta=0.5)$: as $\rho$ crosses one of the threshold values, the system can be found in one of two states, a priori equally probable, characterized by opposite values of $S$. Error bars represent the standard deviation over the simulations run. \label{fig:bifurcation}}
\end{figure*}

Remarkably, this is valid for all the considered configurations: what changes is the number of time steps after which the steady-state, or the consensus, is reached. In particular, the regular lattice (panel a of Figure~\ref{fig:magnetization}) is the configuration where the process is slowest (see Appendix \textbf{A} for further details). As the network is more and more rewired (panels b and c of Figure~\ref{fig:magnetization}), the process becomes faster. Interestingly, further rewiring the network ($\beta>0.5$) does not lead to any appreciable change. Qualitatively speaking, the scale-free configuration (panel d of Figure~\ref{fig:magnetization}) does not show significant differences with respect to the small-world network with $\beta=0.5$; however, the latter reaches the steady-state later, for all the values of $\rho$. It is maybe surprising that the presence of hubs does not speed up the process. However, this apparent paradox could be explained by considering that we are implementing a q-voter model, with an update rule involving only four neighbors at a time: thus, the (potential) influence that hubs could have on large numbers of nodes is drastically reduced. 

Notice also that, for any given configuration, rising $\rho$ shortens the time for reaching the steady-state.

According to Figure~\ref{fig:magnetization} the value $\rho=0.5$ seems to play the role of a threshold, separating two phases of the system: the disordered one, characterized by $M=0$, and the ordered one, with $M$ gradually rising (as a function of $\rho$) until full consensus is reached. As we will show in a while, the behavior of the q-voter model on heterogeneous networks is far richer.

Figure~\ref{fig:phase} shows the value of the magnetization at the steady-state (i.e. after $10^4$ time steps), for two network configurations only (but the same holds true for all the others), as a function of $\rho$. Let us focus on the scale-free configuration (left panel of Figure~\ref{fig:phase}). At a first sight, two distinct phases are visible: the disordered one, characterized by $M=0$ for all the values of $\rho\leq 0.5$, and the ordered one, characterized by a value $M\neq 0$ for $\rho> 0.5$. Thus, the magnetization seems to play the role of the order parameter of a continuous phase transition, while $\rho$ plays the role of control parameter, which can be varied to change the system behavior smoothly. Actually, a closer inspection reveals three different opinion-formation regimes (indicated by different colors), with two distinct threshold values: $\rho^{th}_1\simeq0.45$ separating the flat behavior (in black) from the slowly-rising linear one (in red) and $\rho^{th}_2\simeq0.62$ separating the latter from the rapidly-rising linear one (in green). The insets (zooming on the second transition) reveal that the same qualitative behavior can be observed also for networks with a lower number of agents; what changes is the trend followed by points in the third phase (linear for $N\geq2500$ and quadratic for $N<2500$) with $\rho^{th}_2$ shifting towards lower values ($\simeq0.56$ for $N=500$ agents).

Let us now comment our findings for the Watts-Strogatz configuration (right panel of Figure~\ref{fig:phase}). This time four phases are distinguishable, separated by three threshold values: $\rho^{th}_1\simeq0.55$, $\rho^{th}_2\simeq0.65$ and $\rho^{th}_3\simeq0.7$. However, as the insets reveal, the system loses two of the phases as the number of agents is lowered, showing three linear regimes for $N\geq2500$ and only one quadratic regime for $N<2500$.

However, the analysis of $M$ is somehow limiting because the values of $M$ cannot be negative: this means that the situations where agents reach consensus by adopting the opinions $+1$ and $-1$ are indistinguishable. Thus, we need a quantity able to distinguish the sign of the system final state. To achieve this, we use the \emph{summation of states}
\begin{equation}\label{eq:summation}
S=\frac{\sum_i s_i}{N}=\frac{N_+ - N_-}{N}
\end{equation}
\noindent providing a complementary information with respect to $M$. Plotting the summation $S$ versus the density of conformists, it is possible to achieve further information on the system dynamics. As shown in Figure~\ref{fig:bifurcation}, as the density of conformists rises the system chooses one of two states, characterized by the same absolute value of $S$, but with opposite sign: remarkably, the two states revealed by crossing the thresholds are symmetrically distributed with respect to the horizontal axis. In other words, by rising the density $\rho$ the system is induced to choose one out of two possible states, a priori equally probable, thus breaking its symmetry.

Each point of the phase diagram is the result of an average over $10$ simulations: the obtained values show very small numerical differences, amounting to few percents in the vast majority of cases. When considering the summation of states, to not wash away the information provided by the sign of S, the symmetry-breaking diagram has been obtained by averaging the negative and the positive values separately, maintaining the bi-stable character of the system.
\newline
\newline
\indent The q-voter model shows a very rich behavior, even simply considering agents with two opinions and two characters only, as conformists and non-conformists. Notably, the density of conformist agents $\rho$ strongly affects the consensus reaching process, defining threshold values separating different phases of opinion formation. For $\rho=0$ the two original opinions equally coexist, i.e. 50\% of agents remain in the $+1$ state and 50\% of agents remain in the $-1$ state (with small fluctuations). Then, by progressively rising $\rho$, the system starts showing a magnetization, i.e. a larger number of agents starts sharing the same opinion. This process can be broken down in several phases, suggesting different functional forms of $M(\rho)$, separated by different values of $\rho$. The system crosses the thresholds undergoing a sort of continuous phase transition, as signalled by the magnetization value. 

Apart from the details of the process, the response of the q-voter model to changes of the conformist agent density is remarkably stable across different network topologies: Watts-Strogatz networks with $\beta\geq0.5$ show similar phase-diagrams and symmetry-breaking processes, in turn very similar to the ones observed for the scale-free configuration. The effect of considering heterogeneous topologies is mainly reflected in the speed of the process, which depends on the values of the parameter $\beta$: in particular, the more random the network, the faster the process.

In words, what emerges indicates that different regimes of ``opinion growth'' are identifiable, strongly affected by the density of conformists. Moreover, even if the percentage of conformists drives the society towards a ``prevalent'' opinion (whose diffusion speed grows as more and more conformists are considered) in the case of agents with only two opinions (evenly distributed at $t=0$), the prevalent one cannot be predicted \emph{a priori}.
\newline
\newline
\indent The achieved results open the way to further analyses, as considering agents with more opinions, more and different social traits and other network configurations.

\begin{figure}[t!]
\centering
\includegraphics[width=0.49\textwidth]{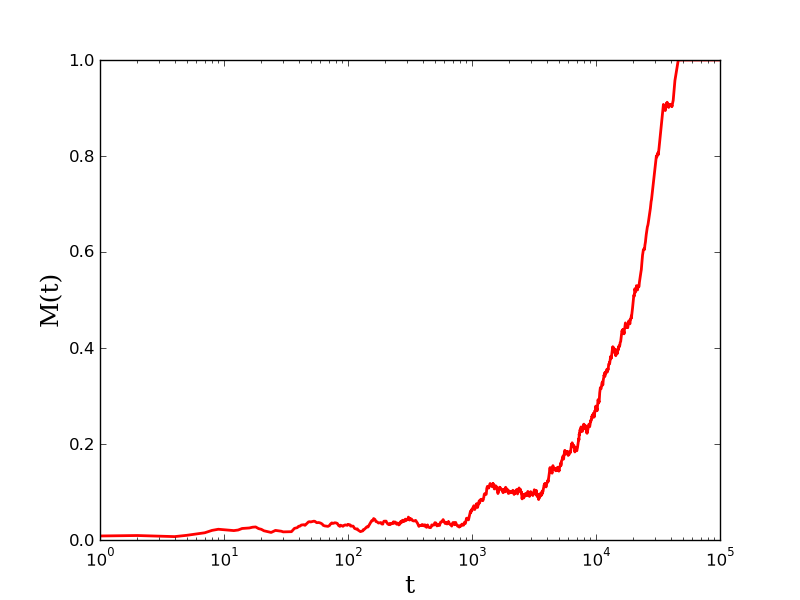}
\caption{\small Evolution of magnetization for $\rho=1$ on the regular ring for $q=4$, up to $10^5$ time steps.\label{fig:long_simulation}}
\end{figure}

\section*{Appendix}

This appendix is devoted to further clarify two important aspects of the q-voter model:

\begin{itemize}
\item[{\bf A:}] how the ordered phase (i.e. consensus) is reached on ring lattices;
\item[{\bf B:}] how the $q$ value affects the results found in the main text.
\end{itemize}

Both issues are investigated by analyzing the magnetization evolution $M(t)$ (averaged over 10 simulations), by considering a population of $N=5000$ agents provided with an average degree $\overline{k}=8$. 

\begin{figure}[t!]
\centering
\includegraphics[width=0.39\textwidth]{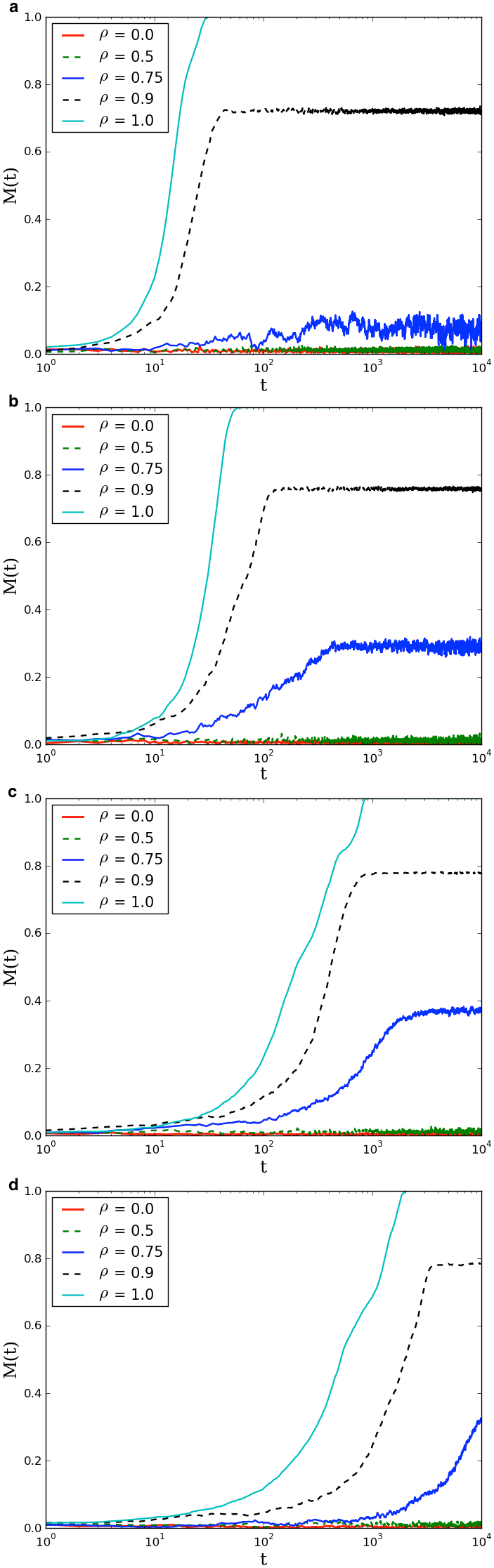}
\caption{\small Evolution of the magnetization in small-world networks ($\beta=0.5$) for different values of $\rho$, for \textbf{a} $q=2$; \textbf{b} $q=3$; \textbf{c} $q=6$; \textbf{d} $q=8$.\label{fig:magnetization_ws_q}}
\end{figure}
\begin{figure}[ht!]
\centering
\includegraphics[width=0.39\textwidth]{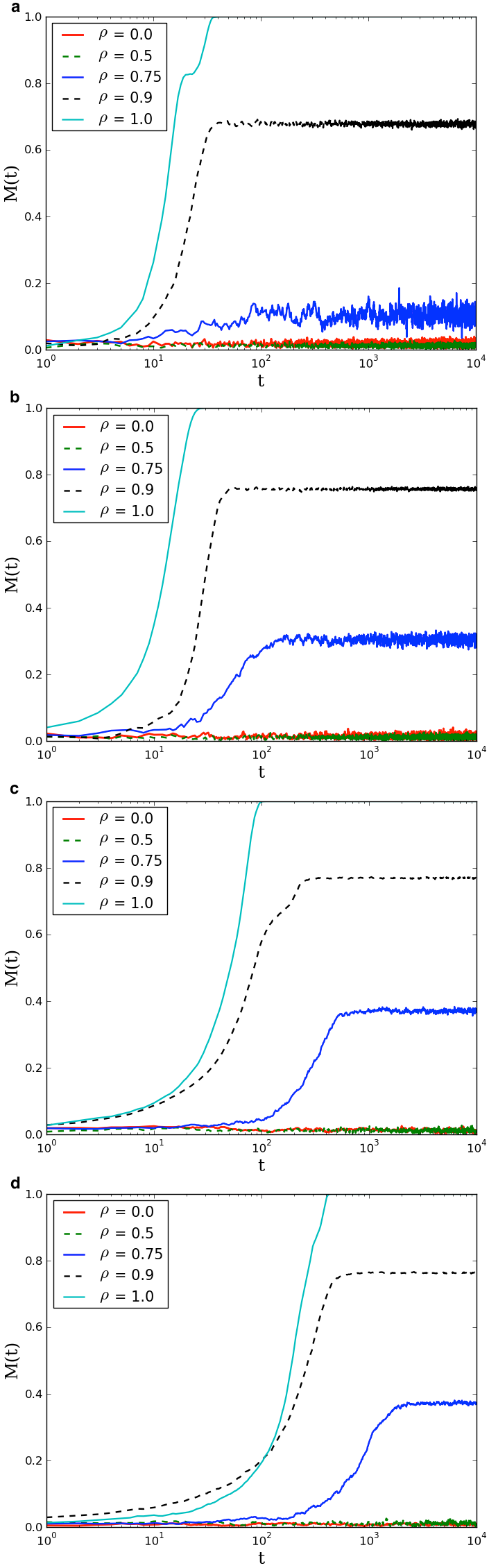}
\caption{\small Evolution of the magnetization in scale-free networks for different values of $\rho$, for \textbf{a} $q=2$; \textbf{b} $q=3$; \textbf{c} $q=6$; \textbf{d} $q=8$.\label{fig:magnetization_sf_q}}
\end{figure}

\subsection{Ordered phase on ring lattices}

As shown in panel b of Figure~\ref{fig:magnetization}, if $\rho=1$ the magnetization of the q-voter model implemented on a ring strongly increases after about $2000$ time steps. However, the value of $M=0.25$ is reached. In order to evaluate whether the population reaches full consensus or a different kind of steady-state, we performed simulations up to $10^5$ time steps. The result is shown in Figure~\ref{fig:long_simulation}: a population composed of conformist agents only (i.e. $\rho=1$) is able to reach the ordered phase we looked for. It is worth to highlight that, this network configuration requires the highest number of time steps to let the agents reach full consensus.

\subsection{Exploring different $q$ values}

We now explore the behavior of the q-voter model by choosing different $q$ values. In particular, we analyze the following range $q\in[2,3,6,8]$. Notice that, if $q=1$, the q-voter model reduces to the classical voter model, as each agent randomly selects one of its neighbors and then assumes the related opinion.

Figures~\ref{fig:magnetization_sf_q} and \ref{fig:magnetization_ws_q} confirm that, qualitatively speaking, the behavior of the q-voter model is not affected by the particular value of $q$, both for the scale-free and the Watts-Strogatz networks. Remarkably, the $\rho$ threshold value above which the system escapes the disordered phase seems to stabilize around $\rho=0.5$, for all the $q$ values (fixing the number of agents at $N=5000$ - see also the discussion in the main text).

As for the value $q=4$ explored in the main text, scale-free networks reach consensus before the Watts-Strogatz ones: for higher values of $q$ the difference can amount to one order of magnitude (see panels d of Figures ~\ref{fig:magnetization_sf_q} and \ref{fig:magnetization_ws_q}).

Moreover, given a particular configuration, the time to reach both the steady state (for $\rho<1$) and the ordered phase ($\rho=1$), increases as $q$ increases. This is intuitive, considering that more time is required to find a higher number of neighbors sharing the same opinion.

Nevertheless, despite the details distinguishing the various simulations, the conclusions drawn for the case $q=4$ can be still generalized to all the considered cases (i.e. to different values of $q$).

\section*{Acknowledgments}
Authors thank Serge Galam for useful suggestions. This work was supported by the Italian PNR project CRISIS-Lab and by Fondazione Banco di Sardegna.


\begin{thebibliography}{99}

\bibitem{econo-socio} Chakrabarti B.K., Chakraborti A., Chatterjee A.: 
Econophysics and Sociophysics: Trends and Perspectives.
\emph{Wiley-VCH Verlag GmbH \& Co.}, Weinheim (2006).

\bibitem{galam01} Galam S.: 
Sociophysics: a review of Galam models.
\emph{International Journal of Modern Physics C}, \textbf{19} (3), 409--440 (2008).

\bibitem{loreto01}
Castellano C., Fortunato S., Loreto V.: 
Statistical physics of social dynamics.
\emph{Rev. Mod. Phys.}, \textbf{81} (2), 591--646 (2009).

\bibitem{perc01} D'Orsogna M.R., Perc M.: 
Statistical physics of crime: A review.
\emph{Physics of Life Reviews}, \textbf{P08013} (2015).

\bibitem{javarone01} Javarone M.A.: 
Network Strategies in Election Campaigns.
\emph{J. Stat. Mech.}, \textbf{P08013} (2014).

\bibitem{galam02} Galam S.: 
Heterogeneous beliefs, segregation, and extremism in the making of public opinions.
\emph{Phys. Rev. E}, \textbf{71} (4), 046123 (2005).

\bibitem{galam03} Galam S.: 
Social paradoxes of majority rule voting and renormalization group.
\emph{J. Stat. Phys.}, \textbf{61} (1990).

\bibitem{galam04} Galam S.: 
Contrarian deterministic effects on opinion dynamics: ``the hung elections scenario''.
\emph{Physica A}, \textbf{333} 453--460 (2004).

\bibitem{intra} Crokidakis N.,  Blanco V.H., Anteneodo C.: 
Impact of contrarians and intransigents in a kinetic model of opinion dynamics.
\emph{arXiv:1401.2880v1} (2014).

\bibitem{javarone02} Javarone M.A.: 
Social Influence in Opinion Dynamics: the Role of Conformity 
\emph{Physica A}, \textbf{414} 19--30 (2014).

\bibitem{aronson01} Aronson E., Wilson T.D. and Akert R.M.: 
Social Psychology. \emph{Pearson Eds.} (2006).

\bibitem{weron04} Nyczka P. , Sznajd-Weron K.: 
Anticonformity or Independence? Insights from Statistical Physics. 
\emph{J. Stat. Phys.}, \textbf{151} 174--202, DOI:10.1007/s10955-013-0701-4 (2013).

\bibitem{weron05} Skorupa B., Sznajd-Weron K., Topolnicki R.: 
Phase diagram for a zero-temperature Glauber dynamics under partially synchronous updates. 
\emph{Phys. Rev. E}, \textbf{86} 051113 (2012).

\bibitem{castellano01} Castellano C., Muoz M.A., Pastor-Satorras R.: 
Non-linear q-voter model.
\emph{Phys. Rev. E}, \textbf{80} 041129 (2009).

\bibitem{galam05} Galam S.: 
Local dynamics vs. social mechanisms: A unifying frame.
\emph{Europhys. Lett.}, \textbf{70} (6), 705--711 (2005).

\bibitem{timpanaro01} Timpanaro A.M., Prado P.C.: 
On the exit probability of the one dimensional q-voter model. Analytical results and simulations for large networks.
\emph{arxiv:1312.2269} (2013).

\bibitem{galam06} Timpanaro A.M., Galam S.: 
An analytical expression for the exit probability of the q-voter model in one dimension.
\emph{arxiv:1408.2734} (2014).

\bibitem{redner01} Liggett T.M.: 
Interacting Particle Systems. \emph{Springer-Verlag}, New York (1985).

\bibitem{weron03} Nyczka P., Sznajd-Weron K., Cislo J.: 
Phase transitions in the q-voter model with two types of stochastic driving. 
\emph{Phys. Rev. E}, \textbf{86} 011105 (2012).

\bibitem{hisakado01} Hisakado M., Mori S.: 
Two kinds of phase transitions in a voting model.
\emph{J. Phys. A}, \textbf{45} 345002, DOI:10.1088/1751-8113/45/34/345002 (2012).

\bibitem{vazquez01} Vazquez F., Eguiluz V.M.: 
Analytical solution of the voter model on uncorrelated networks.
\emph{New J. Phys.}, \textbf{10} 063011 (2008).

\bibitem{sood01} Sood V., Redener S.: 
Voter models on heterogeneous networks.
\emph{Phys. Rev. E}, \textbf{77} 041121 (2008).

\bibitem{satorras01} Moretti, P., Liu S., Castellano C., Pastor-Satorras R.: 
Mean-field analysis of the q-voter model on networks.
\emph{arxiv:1301.7563} (2013).

\bibitem{estrada01} Estrada E.: 
The Structure of Complex Networks: Theory and Applications. \emph{Oxford University Press}, Oxford (2011).

\bibitem{barabasi01} Albert R., Barabasi A.L.: 
Emergence of Scaling in Random Networks.
\emph{Science}, \textbf{286} (5439) 509--512 (1999).

\bibitem{watts01} Watts D.J., Strogatz S.H.: 
Collective dynamics of ``small-world'' networks.
\emph{Nature}, 440--442 (1998).

\bibitem{mobilia01} Mobilia M., Redner S.: 
Majority versus minority dynamics: Phase transition in an interacting two-state spin system.
\emph{Phys. Rev. E}, \textbf{68} (4), 046106 (2003).


\end{thebibliography}
\end{document}